## Original article:

# HAIR HISTOLOGY AS A TOOL FOR FORENSIC IDENTIFICATION OF SOME DOMESTIC ANIMAL SPECIES

Yasser A. Ahmed[1], Safwat Ali[2], Ahmed Ghallab[3*]

[1] Department of Histology, Faculty of Veterinary Medicine, South Valley University, Qena, Egypt
[2] Department of Anatomy and Embryology, Faculty of Veterinary Medicine, Minia University, Minia, Egypt
[3] Department of Forensic Medicine and Toxicology, Faculty of Veterinary Medicine, South Valley University, Qena, Egypt

* Corresponding author: E-mail: ghallab@vet.svu.edu.eg

## ABSTRACT

Animal hair examination at a criminal scene may provide valuable information in forensic investigations. However, local reference databases for animal hair identification are rare. In the present study, we provide differential histological analysis of hair of some domestic animals in Upper Egypt. For this purpose, guard hair of large ruminants (buffalo, camel and cow), small ruminants (sheep and goat), equine (horse and donkey) and canine (dog and cat) were collected and comparative analysis was performed by light microscopy. Based on the hair cuticle scale pattern, type and diameter of the medulla, and the pigmentation, characteristic differential features of each animal species were identified. The cuticle scale pattern was imbricate in all tested animals except in donkey, in which coronal scales were identified. The cuticle scale margin type, shape and the distance in between were characteristic for each animal species. The hair medulla was continuous in most of the tested animal species with the exception of sheep, in which fragmental medulla was detected. The diameter of the hair medulla and the margins differ according to the animal species. Hair shaft pigmentation were not detected in all tested animals with the exception of camel and buffalo, in which granules and streak-like pigmentation were detected. In conclusion, the present study provides a first-step towards preparation of a complete local reference database for animal hair identification that can be used in forensic investigations.

## INTRODUCTION

Animal identification in forensic science is fundamental for many reasons. Analyses of animal remains, e.g. hair or bone, at a criminal scene may help to provide evidence for contact of a suspected assailant (Bertino and Bertino, 2015; Knecht, 2012), or to diagnose some toxic cases e.g. the presence of arsenic, lead or molybdenum in animasl (Harker, 1993; Henderson, 1993; Krumbiegel et al., 2014). In restaurants, animal remain investigation can help to identify meat adulteration, e.g. meats of cat, dog or donkey instead of rabbit, goat or sheep. Furthermore, animal identification is important in case of illegal trade (Cooper and Cooper, 2008; Lawton and Cooper, 2009; Nishant et al., 2017). Moreover, Chernova (2014) provided a proof of principle that hair examination can give some evidence of the age.

Animal species identification can be done based on many features including morphology of animal remains, particularly hair and bone. For example, osteon morphology can be used as a tool to distinguish mammalian from non-mammalian species (Ahmed et al., 2017). Hair morphology is another important tool that can be used to identify animal species (ENFSI, 2015). Four main types of hair were described in different mammals, of which guard hair is the most important in differentiation between various animal species (Knecht, 2012; Tridico, 2005).

The hair consists of two parts, hair root which is embedded in the dermis of the skin, and hair shaft which extends above the epidermis as a cylindrical structure. The hair shaft consists of three distinct morphological layers (Figure 1): medulla (the central layer), cuticle (the outer layer), and the cortex (between the medulla and the cuticle) (Knecht, 2012; Debelica and Thies, 2009; Deedrick and Koch, 2004a). The medulla, the innermost layer of the hair shaft, is a honeycomb-like keratin structure with air spaces in between. The hair medulla can be continuous, discontinuous or fragmental, depending on the species (Deedrick and Koch, 2004a). The cortex contains keratin fibers and pigments which is responsible for the coloration of the hair. The cuticle, the outermost layer, consists of overlapping keratin scales (Deedrick and Koch, 2004a). Two main patterns of cuticle scales were identified: (i) imbricate, this includes ovate, acuminate, elongate, flattened and crenate cuticles; and (ii) coronal, which include simple, serrate or dentate cuticles. The distance between every two successive scale margins can be either close, intermediate or wide, depending on the animal species (Debelica and Thies, 2009). The pattern of the cuticle scales, the type and the diameter of the medulla and/or the characteristics of pigmentation can be used for animal species identification as well as for differentiation between animal and human hair at crime scene (Brunner and Coman, 1974).

The high content of cysteine-containing keratin and dead keratinocytes make the hair resistant to postmortem changes and chemical decomposition (Knecht, 2012; Harkey, 1993). This feature increases the medicolegal importance of hair examination in forensic investigations.

Although several studies and databases are available for human hair examination, complete local databases for comparative animal hair morphology are rarely found. In the present study, we provide a first-step toward a complete reference database for comparative hair morphology of domestic animal species in Upper Egypt.

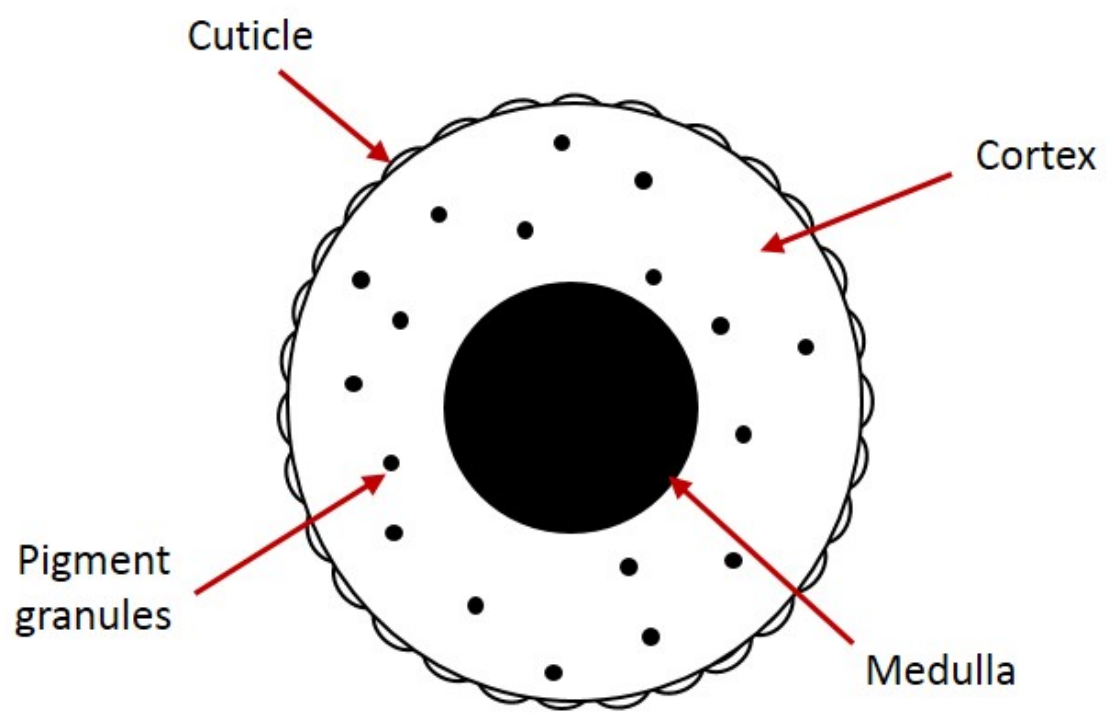

**Figure 1:** Cross section showing the layers of the hair shaft

## MATERIALS AND METHODS

### *Sample collection and preparation*

Dorsal guard hair, between shoulder blades, were taken from three live adult male individuals of domestic buffalo, camel, cow, horse, donkey, sheep, goat, dog, and cat from Luxor and Qena cities, Egypt, in April and May 2018. The collected hair was immersed in 70 % ethanol for 5 minutes in order to remove dirt and sticky non-hairy materials. The middle of the hair shaft, approximately 2 cm length, was cut and prepared for light microscopic examination as described below.

### *Cuticle scale pattern examination*

Cuticle scales were investigated using the gelatin casting method as described by Cornally and Lawton (2016). Briefly, 20 % gelatin was prepared in boiling water. A thin film of gelatin was performed on a clean glass slide. The hair shafts were placed superficially on the gelatin film and left for overnight at room temperature. Subsequently, the hair were removed leaving the imprint of the scales on the gelatin cast. Images were acquired at X40 using a light microscope (Leica DMLS, Germany).

***Medulla and pigmentation examination***

In order to visualize the medulla and pigmentation of the hair, wet-mount or permeant-mount techniques were applied as previously described (Knecht, 2012). Briefly, the collected hair were placed on a glass slide and mounted either by a drop of water (wet-mount) or by DEPX (Thermo Fisher Scientific, Germany) mounting media (permanent mount), and covered with a coverslip. Images were acquired at X100 using a light microscope (Leica DMLS, Germany).

***Result evaluation***

Cuticle scale patterns, types of medulla, and pigmentation features were identified and compared in different animal species using the available animal hair keys in the literature (Knecht, 2012; Debelica and Thies, 2009; Brunner and Coman, 1974; Cornally and Lawton, 2016; Huffman and Wallace, 2012).

## RESULTS

Based on the hair cuticle scale patterns, type and diameter of the medulla and the hair pigmentation domestic animals (large ruminants, equine, small ruminants, and canine) were identified and compared (Table 1; Figures 2 and 3).

***Comparative analysis of hair cuticle scales for identification of domestic animals***

The hair cuticle scale patterns, the margin type, shape and distance between scales were visualized and compared in the tested animals (Table 1; Figure 2). The cuticle scales were imbricate in all tested animals except in donkey, in which coronal scales were identified. Based on scale margin type, shape and distance the tested animals were clearly differentiated. In buffalo, rippled cuticle margins, close-distant, double-chevron shaped were detected. Whereas, cuticle scales with smooth margins, irregular mosaic shape, and wide-distant were found in sheep. In dog, the scale margin type and distance were similar to that of sheep, but regular in shape. In cat, crenate scale margins, irregular in shape with close distances were detected. In horse and camel, the scale margin type, shape and distance were similar, appeared as crenate irregular waves with intermediate distances. In cow, the scale margins were also crenate with intermediate distances, but appeared as regular waves in shape (Table 1; Figure 2).

***Comparative analysis of hair medulla and pigmentation for identification of domestic animals***

In addition to the hair scale morphology, the type, size and margins of the medulla as well as the pigmentation were used to identify and compare the tested domestic animals (Table 1; Figure 3). All tested animals showed continuous type of medulla with the exception of sheep, in which the medulla was fragmental. In buffalo and donkey, the diameter of the medulla was very wide covering almost the entire hair shaft; the margins were smooth. In contrast, in horse the diameter of the medulla was very narrow, less than one third of the hair shaft, and the margins were serrated or notched. In goat, the diameter of the medulla was slightly wider than in horse, covering approximately one third of the hair shaft, with serrated margins. Approximately 50 % of the hair shaft was covered by the medulla in dog, and the margins were smooth. In camel, cow and cat, the medulla occupied approximately two thirds of the hair shaft; the margins were smooth in camel and cow, but slightly serrated in cat. Hair shaft pigmentation were not detected in all tested animals with the exception of camel and buffalo, in which granules and streak-like pigmentation were detected (Table 1; Figure 3).

**Table 1:** Comparative analysis of hair morphology in different animal species

| Species | | Cuticle scales | | | | Medulla | | | Pigmentation |
|---|---|---|---|---|---|---|---|---|---|
| | | **Pattern** | **Margin type** | **Margin Shape** | **Margin distance** | **Type** | **Diameter** | **Margin** | |
| **Large ruminants** | **Buffalo** | Imbricate | Rippled | Double Chevron | Close | Continuous | Almost entire shaft | smooth | Granules and streaks-like pigments |
| | **Camel** | Imbricate | Crenate | Irregular wave | Intermediate | Continuous | More than 1/2 | smooth | Granules and streaks-like pigments |
| | **Cow** | Imbricate | Crenate | Regular wave | Intermediate | Continuous | More than 1/2 | smooth | No pigmentation |
| **Equine** | **Horse** | Imbricate | Crenate/ smooth | Irregular wave | Intermediate | Continuous | Less than 1/3 | Serrated or notched | No pigmentation |
| | **Donkey** | Coronal | Crenate | Regular wave | Intermediate | Continuous | Almost entire shaft | smooth | No pigmentation |
| **Small ruminants** | **Sheep** | Imbricate | Smooth | Irregular mosaic | Wide | Fragmental | - | - | No pigmentation |
| | **Goat** | Imbricate | Crenate | Regular wave | Intermediate | Continuous | About 1/3 | serrated | No pigmentation |
| **Canine** | **Dog** | Imbricate | Smooth | Regular petal | Wide | Continuous | About 1/2 | smooth | No pigmentation |
| | **Cat** | Imbricate | Crenate | Irregular wave | Close | Continuous | More than 1/2 | Slightly serrated | No pigmentation |

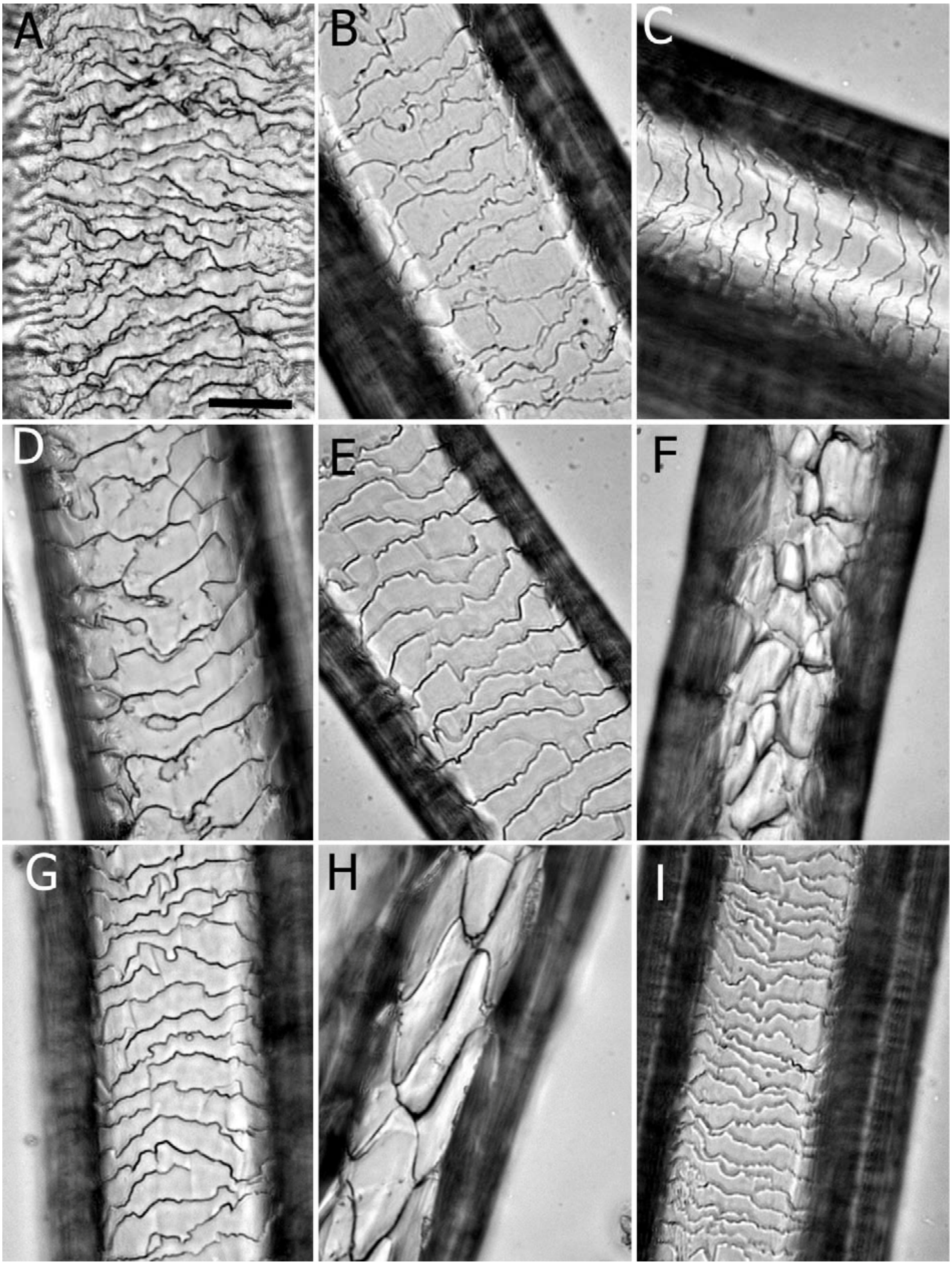

**Figure 2:** Morphology of the hair cuticle scales in different animal species. **A**. buffalo; **B**. camel; **C**. cow; **D**. horse; **E**. donkey; **F**. sheep; **G**. goat; **H**. dog; **I**. cat. Scale bar: 25 µm

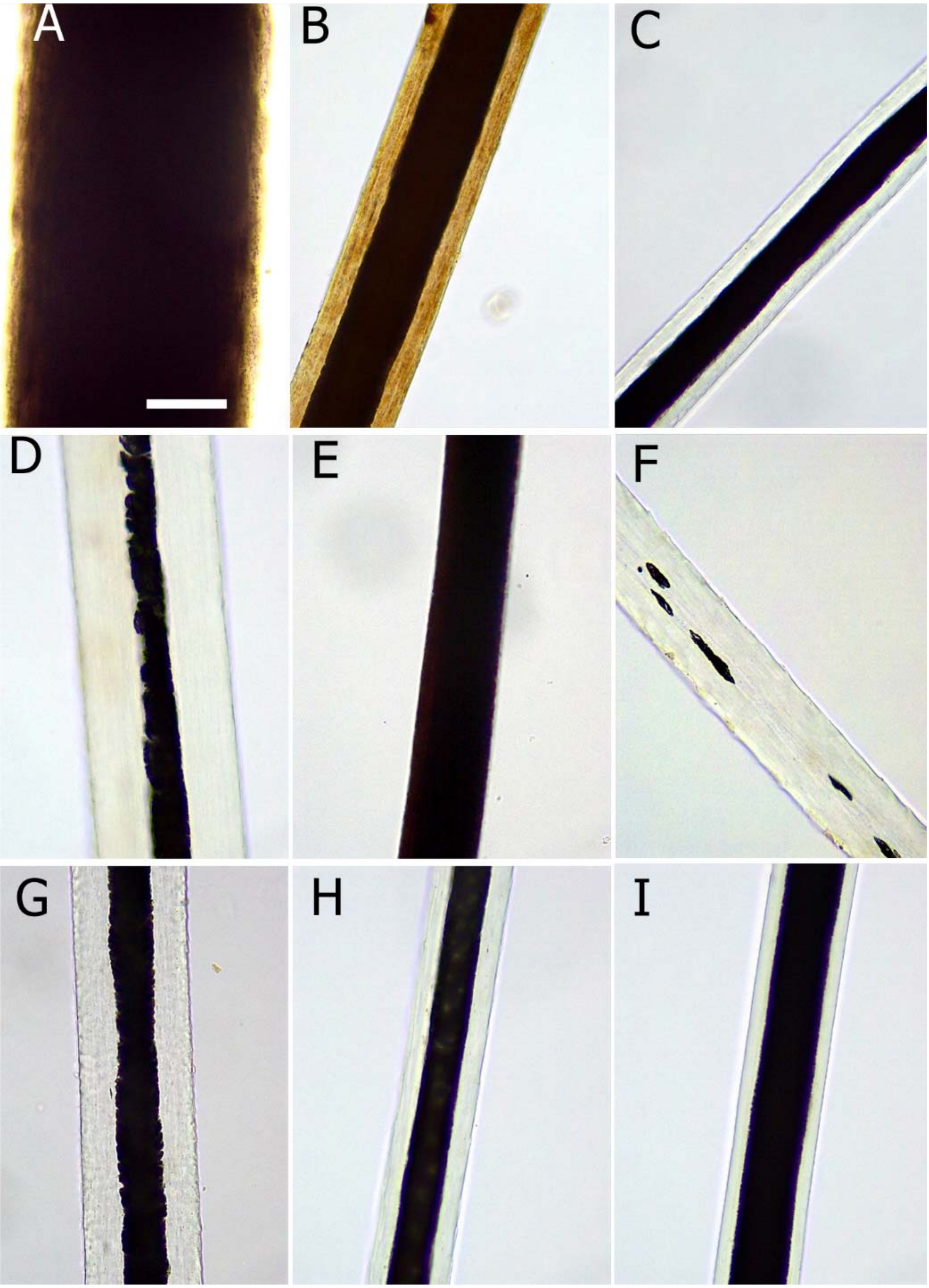

**Figure 3:** Morphology of the hair medulla in different animal species. **A**. buffalo; **B**. camel; **C**. cow; **D**. horse; **E**. donkey; **F**. sheep; **G**. goat; **H**. dog; **I**. cat. Scale bar: 62.5 µm

## DISCUSSION

Hair examination plays an important role in forensic investigations. Analysis of hair at a crime scene might be helpful to trace a suspicious contact. Moreover, it may help to diagnose some toxic cases, e.g. arsenic, lead and molybdenum toxicity, and identify the duration of exposure (Harkey, 1993; Henderson, 1993; Krumbiegel et al., 2014). In human, hair examination is used also to detect drug abuse e.g. heroin, amphetamine and cannabis (Sen, 2010). Resistance of the hair to postmortem changes and chemical decomposition due to the high content of cysteine-containing keratin (Knecht, 2012; Harkey, 1993), added to its value in forensic investigations. However, the field of veterinary forensic medicine in general, including hair examination, is not well-developed in comparison to human forensics. Although few local keys for animal hair identification are available worldwide, e.g. in Europe (Keller, 1978, 1980; Lochte, 1938) and United States (Mayer, 1952), such regional keys are lacking in Egypt. In the present study, comparative histological analysis of domestic animal hair was performed, including large ruminants (buffalo, camel and cow), small ruminants (sheep and goat), equine (horse and donkey), and canine (dog and cat). Based on the morphology of the hair cuticle scales, medulla and pigmentation, comparative characteristic features were recorded, using the currently available animal hair keys in literature as a guide (Knecht, 2012; Debelica and Thies, 2009; Brunner and Coman, 1974; Cornally and Lawton, 2016; Huffman and Wallace, 2012).

Our results are similar to the available literature (Knecht, 2012: Debelica and Thies, 2009; Brunner and Coman, 1974; Cornally and Lawton, 2016; Huffman and Wallace, 2012; Mukherjee et al., 2016; Trivedi, 2015), but with few exceptions. The present analysis revealed that the hair cuticles in cow are of imprecate type, the margins are crenate, regular waved and intermediate-distant. The medulla is continuous, occupying more than half of the hair shaft, and no pigments were seen in the cortex. Trivedi (2015) reported similar results with the exception that the scale margins are wide-distant, partially smooth and partially notched. In equine, our results revealed that the medulla is continues, occupying less than one third of the shaft in donkey and cover almost the entire shaft in horse. However, Gharu and Trivedi (2015), showed that the hair medulla in equine is fragmental. The hair scales in sheep are imbricate, smooth, irregular mosaic and wide-distant; the medulla is fragmental. In goat, the cuticle scale pattern is imbricate with crenate, irregular-wave, intermediate-distant margins; the medulla is continuous occupying approximately one third of the hair shaft. This result is not in agreement with reports in the literature describing that both in sheep and goat the cuticle scales are arranged irregularly, and the medulla is continuous. The hair cuticle scales in dog is imbricate, smooth, petal and wide-distant, and the medulla is continuous occupying approximately half of the hair shaft; this is different from the vacuolated structure of the hair medulla described by Mukherjee et al. (2016). A possible explanation for this discrepancy could be the adaptive changes due to different climatic conditions.

It is important to differentiate between human and animal hair in forensic investigations. This can easily be done by inspection of the hair medulla. In most of the tested animals the medulla is continuous with wide diameter, whereas, in human the hair medulla is usually very thin, fragmental or may be absent in some cases (Deedrick and Koch, 2004b). An exception based on the present results is sheep, in which also a thin fragmental hair medulla was detected. Therefore, careful analysis of hair scales and pigmentation is essential to differentiate between hair of human and sheep.

While the present results gives a clear comparative picture between the tested animal species, one limitation is that it is difficult to differentiate between various animal strains within the same family based only on hair morphology. For this purpose, further analyses are required.

In conclusion, the present study provides a first-step towards preparation of a local reference database for animal hair identification that can be used in forensic investigations. Further studies are required using more sophisticated techniques to have a complete local atlas for animal identification.